\documentclass[3p,number,sort&compress]{elsarticle}
\journal{Computer-Aided Design}

\usepackage[charter]{mathdesign}

\usepackage[english]{babel}
\usepackage[T1]{fontenc}
\usepackage[utf8x]{inputenc}
\usepackage{microtype}
\usepackage{amsmath}
\usepackage{array}
\usepackage{graphicx}
\usepackage{indentfirst}
\usepackage{booktabs}
\usepackage{hyperref}
\usepackage{pdfpages}
\usepackage{float}
\usepackage{tikz-qtree}
\usepackage{flafter}
\usepackage{siunitx}
\usepackage{enumitem}
\usepackage{algorithm}
\usepackage{algpseudocode}
\usepackage[capitalise,nameinlink]{cleveref}
\usepackage[colorinlistoftodos]{todonotes}

\DeclareMathOperator{\sign}{sign}
\newcolumntype{L}[1]{>{\raggedright\let\newline\\\arraybackslash\hspace{0pt}}m{#1}}
\newcolumntype{C}[1]{>{\centering\let\newline\\\arraybackslash\hspace{0pt}}m{#1}}
\newcolumntype{R}[1]{>{\raggedleft\let\newline\\\arraybackslash\hspace{0pt}}m{#1}}

\begin{document}

\title{Point containment algorithms for constructive solid geometry with unbounded primitives}

\author[anl]{Paul K. Romano\corref{cor1}}
\ead{promano@anl.gov}
\cortext[cor1]{Corresponding author. Tel.: +1 630 252 6779.}

\author[um]{Patrick A. Myers}
\ead{myerspat@umich.edu}

\author[ornl]{Seth R. Johnson}
\ead{johnsonsr@ornl.gov}

\author[uned]{Alja\u{z} Kol\u{s}ek}
\ead{akolsek@ind.uned.es}

\author[anl]{Patrick C. Shriwise}
\ead{pshriwise@anl.gov}

\address[anl]{Argonne National Laboratory, 9700 S. Cass Ave, Lemont, IL 60439, United States}
\address[um]{University of Michigan, Ann Arbor, MI 48103, United States}
\address[ornl]{Oak Ridge National Laboratory, Oak Ridge, TN 37830, United States}
\address[uned]{Universided Nacional de Educaci\'{o}n a Distancia, Madrid, Spain}

%-------------------------------------------------------------------------------
%   STYLE OPTIONS
%-------------------------------------------------------------------------------

%-------------------------------------------------------------------------------

\begin{abstract}
We present several algorithms for evaluating point containment in constructive
solid geometry (CSG) trees with unbounded primitives. Three algorithms are
presented based on postfix, prefix, and infix notations of the CSG binary
expression tree. We show that prefix and infix notations enable short-circuiting
logic, which reduces the number of primitives that must be checked
during point containment. To evaluate the performance of the algorithms, each
algorithm was implemented in the OpenMC Monte Carlo particle transport code,
which relies on CSG to represent solid bodies through which subatomic particles
travel. Two sets of tests were carried out. First, the execution time to
generate a high-resolution rasterized image of a 2D slice of a detailed CSG
model of the ITER tokamak was measured. Use of both prefix and infix notations
offered significant speedup over the postfix notation that has traditionally
been used in particle transport codes, with infix resulting in a 6$\times$
reduction in execution time relative to postfix. We then measured the execution
time of neutron transport simulations of the same ITER model using each of the
algorithms. The results and performance improvements reveal the same trends as
for the rasterization test, with a 4.59$\times$ overall speedup using the infix
notation relative to the original postfix notation in OpenMC.
\end{abstract}


\maketitle

%-------------------------------------------------------------------------------
\section{Introduction}

% Lead first paragraph by talking about point containment?

\begin{figure}
    \centering
    \begin{tikzpicture}
    [level distance=10mm, every node/.style={minimum width=2em,circle,draw},
    edge from parent/.style={draw,edge from parent path={(\tikzparentnode) -- (\tikzchildnode)}},
    level 1/.style={sibling distance=10mm},
    level 2/.style={sibling distance=10mm}]
    \Tree
    [.$\cup$
        [.$\cap$
        \edge[]; {$A$}
        \edge[]; {$B$}
        ]
        [.$\cap$
        \edge[]; {$C$}
        \edge[]; {$D$}
        ]
    ]
    \end{tikzpicture}
    \centering
    \caption{A CSG binary expression tree constructed with primitives ($A$, $B$, $C$, and $D$) and set intersection ($\cap$) and union ($\cup$) operators.}
    \label{fig:tree}
\end{figure}
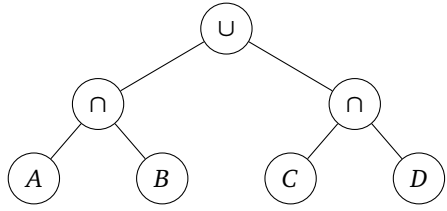

Constructive solid geometry (CSG) is a technique for representing rigid solids
using Boolean set operations applied to simple geometric
primitives~\citep{CSG,CSG2,requicha1980acmcs}. In CSG, a solid is represented as
an ordered, binary tree where the non-leaf nodes are Boolean set operators
(union, intersection, difference) and the leaf nodes represent geometric
primitives. An example of such a CSG tree is illustrated in \cref{fig:tree}. CSG
representations can be based on bounded primitives (cuboids, spheres, cylinders,
cones, etc.) or unbounded primitives using general half-spaces.

CSG has applications in geometric modeling, computer-aided design, computer
graphics, computational science, and other areas where accurate spatial modeling
is required~\citep{CSG,CSG2,Boolean}. In particular, CSG is widely used for
geometric modeling in Monte Carlo (MC) particle transport simulations through
codes such as OpenMC~\citep{romano2015ane}, MCNP~\citep{kulesza2022lanl},
Shift~\citep{pandya2016jcp}, FLUKA~\citep{bohlen2014nds}, and
Serpent~\citep{leppanen2015ane}. MC particle transport itself has applications
across a wide range of scientific domains, such as nuclear reactor design,
radiation shielding, medical physics, high-energy experimental physics, and
fusion energy.

Most MC particle transport codes---including OpenMC, which was used for the
present study---use a CSG representation based on unbounded primitives defined
by inequalities of implicit surfaces. An implicit surface is a surface in
Euclidean space defined by an equation $f(\mathbf{r})$ = 0. By convention, the
inside of the surface is defined as the set of points for which
$f(\mathbf{r})<0$ (the negative half-space) and the outside of the surface is
defined by $f(\mathbf{r})>0$ (the positive half-space). Thus, the leaf nodes in
the CSG tree are implicit surface inequalities. Surface inequalities are then
combined with Boolean set operations to create more complex solids. MC particle
transport codes are typically limited to primitives based on algebraic surfaces
(planes, quadrics, and sometimes torii); nearly all allow the use of set
intersection and union operators, while some also provide set difference or
complement operators.

Although the overall runtime of a MC particle transport simulation is
distributed across many different operations, the time spent on geometric
operations can be significant for large, geometrically complex models. One
important geometric operation is \emph{point containment}---given a point
$\mathbf{r}$ in space, determine whether it is contained within a particular
solid. In MC particle transport simulations, this operation is used extensively
for determining a particle's location as it traverses through the geometry. In
this work, we explore several algorithms for evaluating point containment in CSG
trees with unbounded primitives with a goal of ultimately minimizing the
execution time of MC particle transport simulations.

\subsection{Point Containment}

\noindent Evaluating point containment for a single unbounded primitive based on
an implicit surface is straightforward. First, the implicit surface equation is
evaluated at the point, $f(\mathbf{r})$, and the sign of the result is compared
against the specified sign of the primitive in the CSG expression. If the sign
matches, the point is contained within the primitive. Extending this to a
general solid defined by a CSG tree, we can evaluate point containment by first
checking point containment for all primitives, yielding a binary expression tree
with Boolean leaf nodes, i.e., a Boolean expression. Evaluating this Boolean
expression indicates whether the point is contained.

While evaluating point containment using a CSG tree is straightforward in
principle, an important question concerning implementation remains: how should
the CSG tree be represented in memory? In all the algorithms we explore here,
the CSG tree is flattened into a one-dimensional data structure to improve
locality of reference and minimize memory use. With a flattened data structure,
the tree can be represented using either an infix, postfix, or prefix notation.
Taking the CSG tree from \cref{fig:tree} as an example, the three notations are
as follows:
\begin{itemize}
    \item Infix: $(A \cap B) \cup (C \cap D)$
    \item Postfix: $A\:B\cap C\:D\cap\cup$
    \item Prefix: $\cup\cap A\:B\cap C\:D$
\end{itemize}
While infix notation is commonly used in written language, postfix and prefix
notations have the advantage of 1) being easily parsed by a computer, 2)
simplifying traversal of the tree, and 3) not requiring parentheses to enforce
precedence of operators.

In addition to the question of which notation is to be preferred for
representing a CSG tree, a related question is whether it is possible to take
advantage of short-circuiting evaluation for point containment queries, also
sometimes referred to as ``early-out''~\citep{jansen1991acmtog}. That is, when an
intersection operator is encountered in the binary expression tree, if the first
operand is found to be false it is not necessary to evaluate the second operand
at all. Similarly with a union operator, the first operand evaluating to true
obviates the need to evaluate the second operand. When implicit surface
inequalities are used as primitives in the CSG tree, evaluating the operand is
equivalent to evaluating an implicit surface function $f(\mathbf{r})$, which may
bear a non-trivial cost. Thus, any notation that enables the use of
short-circuit evaluation can help minimize the number of function evaluations.

For a given notation, evaluating the CSG binary expression may also require the
use of a stack to manage the intermediate results of point containment on the
given primitives. For postfix notation, several methods have been discussed in
the literature, including bit-sequential and bit-parallel
methods~\citep{jansen1991acmtog}.

To our knowledge, there are no existing works in the literature discussing the
use of infix, postfix, and prefix notation for representing CSG trees as used in
MC particle transport codes. Based on our own survey of several popular codes,
we have found that the OpenMC code through version 0.13.1~\citep{openmc-0131},
Serpent, and Shift rely on a postfix notation for storing CSG trees, whereas
MCNP relies on an infix notation. We are not aware of any codes that rely on
prefix notation for storing CSG trees. Additionally, none of the codes surveyed
use short-circuit evaluation for point containment queries as far as we are
aware\footnote{As discussed further in \cref{sec:results}, some
codes---including OpenMC---do implement short-circuit evaluation for CSG
expressions involving only set intersection operators. However, to our knowledge
none use short-circuit evaluation for \emph{all} expressions.}.

\subsection{Summary of Contributions}

\noindent The first contribution of this work is to explicitly sketch out
algorithms for evaluating point containment on CSG trees using postfix notation
(and postorder traversal) as well as algorithms for evaluation using infix and
prefix notations. To our knowledge, no such descriptions for infix and prefix
notations have been described in the existing literature. As will be shown, both
infix and prefix notations enable the use of short-circuit evaluation; our
second contribution is---as part of the algorithm descriptions---to demonstrate
how short-circuit evaluation can be implemented when using infix or prefix
notation.

For postfix and prefix notations, postorder and preorder traversal of the CSG
tree require that a stack be maintained: a stack of Boolean values for postfix
notation and a stack of operators for prefix notation. We discuss several
strategies for minimizing the memory footprint of data structures for managing
the stack and the associated computational expense of managing the stack.

The algorithm descriptions themselves are useful for understanding \emph{how}
point containment queries can be performed using various notations, but they do
not by themselves provide any hard evidence why one or the other should be
preferred. To study the performance of these algorithms, we have implemented all
of them in the OpenMC particle transport code and performed simulations of the
ITER fusion experiment, relying on an extremely detailed CSG model called
E-lite~\citep{juarez2021nature}. These simulations demonstrate that, for problems
that involve complex CSG trees, both prefix and infix notations offer
substantial performance improvement by enabling short-circuit evaluation. For
postfix and prefix notation, we also study how the choice of a stack data
structure impacts performance.

\subsection{Outline of Paper}

\noindent \Cref{sec:methodology} describes the algorithms for CSG point
containment queries based on the three expression notations: postfix, prefix,
and infix. It also describes data structures for managing operand/operator
stacks. \Cref{sec:results} describes performance results of the various
algorithms as implemented in MC particle transport simulations based on the ITER
experiment. Finally, \cref{sec:conclusions} provides overall conclusions based
on the results and describes the limits of the present study.

%-------------------------------------------------------------------------------
\section{Methodology}
\label{sec:methodology}

\subsection{Postfix Evaluation}

\begin{figure}[ht]
  \centering
  \includegraphics[width=1.5in]{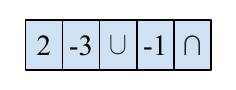}
  \caption{Memory representation (left-to-right) of the postfix notation for the
  expression ``${-1} \cap ({+2} \cup {-3})$''.}
  \label{fig:postfix-memory}
\end{figure}

\noindent As mentioned before, many MC particle transport codes rely on postfix
notation to perform evaluation, owing to the ease by which it enables one to
perform a postorder traversal (scanning the expression left to right). Most
often, the user input for CSG solids in MC particle transport codes is provided
using infix notation, which is then converted into a postfix notation using
Dijkstra's shunting yard algorithm~\citep{Dijkstra}. The postfix notation, $T$,
is stored in memory as a flattened array. While there are many possible
approaches for representing the notation in memory, we outline the approach
taken in OpenMC, which we believe to be representative of other MC particle
transport codes as well. Binary CSG expression trees in OpenMC have an input
representation based on the following conventions. First, implicit surfaces are
referenced through a unique, positive integer. Implicit surface inequalities are
represented by referring to the associated unique, positive integer identifying
the implicit surface along with the unary positive and negative operators, which
refer to the negative and positive half-spaces of the implicit surface,
respectively. Thus, given two implicit surfaces identified by the integers 1 and
2, the expression ``${-1} \cap {-2}$'' would represent the intersection of their
negative half-spaces and ``${-1} \cup {-2}$'' would represent the union of their
negative half-spaces. With these conventions in mind, \cref{fig:postfix-memory}
shows an example of how the postfix notation for an expression would be
represented in memory. It suffices to store each element of the array as a
signed 32-bit integer. The largest integer values (e.g., $2^{31}-1$) are
reserved for operators.

\cref{alg:postfix} outlines the procedure for evaluating point containment based
on an expression represented using postfix notation. This procedure is
effectively performing a depth-first traversal of the original binary expression
tree. Each time a primitive is encountered, the $\textsc{Evaluate}$ function,
outlined in \cref{alg:evaluate}, is called. Anytime a set intersection operator
appears in the original binary expression tree, a logical AND operator is used
on the results of the $\textsc{Evaluate}$ function, which returns a Boolean
value. Similarly, a logical OR operator is used in place of the set union
operator. This algorithm requires a full traversal of the expression because
every evaluation of an operator requires that its operands (the implicit surface
inequalities) be evaluated first. Additionally, a stack of Boolean values
(evaluated operands) is necessary to keep track of the results of evaluating the
primitives.
\begin{algorithm}
  \caption{Postfix evaluation algorithm for point containment.}
  \label{alg:postfix}
  \begin{algorithmic}
    \Require $T \equiv$ CSG expression in postfix notation
    \Require $\mathbf{r} \equiv$ point in Euclidean space \\
    \Function{ContainsPostfix}{$T,\mathbf{r}$}
    \State $S \equiv$ empty stack for Boolean values
    \For{$i \gets 1,$ \textsc{Size($T$)}}
      \State{$x \gets T[i]$} \Comment{Get next node}
      \If{$x$ is a primitive}
        \State $b \gets$ \textsc{Evaluate}($x,\mathbf{r}$)
        \State \textsc{Push($S$, $b$)} \Comment{Store primitive containment}
      \Else
        \State $b_1 \gets$ \textsc{Pop}($S$) \Comment{Pop two operands}
        \State $b_2 \gets$ \textsc{Pop}($S$)
        \If{$x$ is $\cup$}
          \State \textsc{Push}($S, b_1 \lor b_2$) \Comment{Apply logical OR}
        \ElsIf{$x$ is $\cap$}
          \State \textsc{Push}($S, b_1 \land b_2$) \Comment{Apply logical AND}
        \EndIf
      \EndIf
    \EndFor
    \State \textbf{return} \textsc{Pop}($S$) \Comment{Return top of stack}
    \EndFunction
  \end{algorithmic}
\end{algorithm}

Also note that \cref{alg:postfix} assumes the expression only contains set union
and intersection operators. Set difference and complement operators can be
removed through the use of De Morgan's laws prior to application of this
procedure.

\begin{algorithm}
  \caption{Evaluating an implicit surface inequality.}
  \label{alg:evaluate}
  \begin{algorithmic}
  \Require $x \equiv$ node representing the implicit surface inequality
  $f(\mathbf{r}) > 0$ or $f(\mathbf{r}) < 0$. Let $x.\text{sign}$ be 1 when
  $f(\mathbf{r})$ is greater than zero and -1 when it is less. \\
  \Require $\mathbf{r} \equiv$ point in Euclidean space \\

    \Function{Evaluate}{$x,\mathbf{r}$}
    \State $f_x \gets$ function $f(\mathbf{r})$ associated with node $x$
    \State $\phi \gets f_x(\mathbf{r})$
    \State \textbf{return} $\sign(\phi) = x.\text{sign}$
    \EndFunction
  \end{algorithmic}
\end{algorithm}

\subsection{Prefix Evaluation}

\begin{figure}[ht]
  \centering
  \includegraphics[width=1.5in]{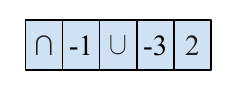}
  \caption{Memory representation (left-to-right) of the prefix notation for the
  expression ``${-1} \cap ({+2} \cup {-3})$''.}
  \label{fig:prefix-memory}
\end{figure}

\noindent While prefix evaluation has not been previously used on binary
expression trees for CSG point containment operations to our knowledge, it has
an advantage over postfix evaluation in that it enables the use of
short-circuiting logic and would thus avoid unnecessary evaluation of implicit
surface inequalities. After a CSG expression has been converted to a postfix
notation using the shunting yard algorithm, obtaining the prefix notation is as
simple as reversing the postfix notation; an example prefix notation is shown in
\cref{fig:prefix-memory}. Evaluation of the prefix notation then proceeds as in
\cref{alg:prefix}. In this algorithm, \emph{set operators} are pushed onto an
operator stack rather than Boolean values as in postfix evaluation. By the time
the second operand of a set operator is encountered, the first operand has
already been evaluated, which makes it straightforward to short circuit. The
algorithm for short circuiting is shown in \cref{alg:short-circuit} and relies
on counting the number of operators and primitives encountered. As before,
\cref{alg:prefix} assumes that only set union and intersection operators are
used; other operators are assumed to have been eliminated via De Morgan's laws.
\begin{algorithm}
  \caption{Prefix evaluation algorithm with short circuiting}
  \label{alg:prefix}
  \begin{algorithmic}
    \Require $T \equiv$ CSG expression in prefix notation
    \Require $\mathbf{r} \equiv$ point in Euclidean space \\
    \Function{ContainsPrefix}{$T,\mathbf{r}$}
    \State $S \equiv$ empty stack for operators
    \State $b \gets \text{FALSE}$ \Comment{Boolean value to be returned}
    \State $i \gets 1$ \Comment{Index in $T$}
    \While{$i <$ \textsc{Size($T$)}}
      \State{$x \gets T[i]$} \Comment{Get next node}
      \If{$x$ is a primitive}
        \State $b \gets$ \textsc{Evaluate}($x,\mathbf{r}$)
        \While{\textsc{Size}($S$) $> 0$}
          \State $y \gets $\textsc{Pop}($S$) \Comment{Get node from stack}
          \If{$y$ is $\cup$ \textbf{and} $b = \text{TRUE}$}
            \State \textsc{ShortCircuit}($T$, $i$)
          \ElsIf{$y$ is $\cap$ \textbf{and} $b = \text{FALSE}$}
            \State \textsc{ShortCircuit}($T$, $i$)
          \Else
              \State \textbf{break} \Comment{Status value is $b$}
          \EndIf
        \EndWhile
      \Else
        \State \textsc{Push}($S, x$) \Comment{Store operator on stack}
      \EndIf
      \State $i \gets i + 1$
    \EndWhile
    \State \textbf{return} $b$
    \EndFunction
  \end{algorithmic}
\end{algorithm}

\begin{algorithm}
  \caption{Skip part of prefix expression for short-circuiting.}
  \label{alg:short-circuit}
  \begin{algorithmic}
    \Require $T \equiv$ CSG expression in prefix notation
    \Require $i \equiv$ Index in $T$ \\
    \Function{ShortCircuit}{$T,i$}
      \State $N_\text{operator} \gets 0$ \Comment{Number of operators}
      \State $N_\text{primitive} \gets 0$ \Comment{Number of primitives}
      \While{$N_\text{primitive} \ne N_\text{operator} + 1$}
        \State $i \gets i + 1$
        \State $x \gets T[i]$ \Comment{Get next node}
        \If{$x$ is a primitive}
          \State $N_\text{primitive} \gets N_\text{primitive} + 1$
        \Else
          \State $N_\text{operator} \gets N_\text{operator} + 1$
        \EndIf
      \EndWhile
    \EndFunction
  \end{algorithmic}
\end{algorithm}

\subsection{Stack Memory Management}

\noindent Both postfix (\cref{alg:postfix}) and prefix (\cref{alg:prefix})
evaluation algorithms require that a stack of objects be maintained. A stack
data structure in C++ is typically managed through the \texttt{std::vector}
class from the Standard Template Library (STL), which encapsulates a dynamically
sized array. As the size of the stack grows, memory may be dynamically
allocated, which is undesirable for a number of reasons. First, point
containment queries in a MC particle transport simulation occur from multiple
threads; because dynamic memory allocation may utilize mutual exclusion locks to
ensure thread-safety, it can significantly degrade threaded performance. Second,
on some compute architectures, such as GPUs, dynamic memory allocation may not
be possible at all during the execution of a kernel. Here, we present techniques
for minimizing memory requirements of the stack data structures utilized in the
postfix and prefix evaluation algorithms such that dynamic memory allocation can
be almost entirely avoided.

For postfix evaluation, a stack of Boolean values must be maintained. While the
STL has a specialization of \texttt{std::vector} for the \texttt{bool} datatype
that is nominally space-efficient, the actual behavior is implementation-defined
and complicated by the fact that the \texttt{bool} datatype in C++ has a size of
at least 1 byte (rather than 1 bit). Consequently, the implementation of
\texttt{std::vector<bool>} often requires 8 bits per Boolean value. By avoiding
the \texttt{std::vector} altogether, it is possible to achieve true bit-packing
whereby each Boolean value is stored using a single bit. With a data structure
containing a single 32-bit unsigned integer that stores the bit-packed Boolean
values, one can carry out \cref{alg:postfix} with as many as 32 Boolean values
on the stack at any given point. Efficient stack operations (push, pop, top,
indexing) can be achieved through the use of bitwise operators applied to the
underlying integer. While a maximum stack size of 32 may seem limiting, we
observed that for the ITER E-lite model (discussed at length in
\cref{sec:results}), which has a high degree of geometric complexity, the stack
size was never greater than 9.

\begin{figure}[ht]
  \centering
  \includegraphics[width=3in]{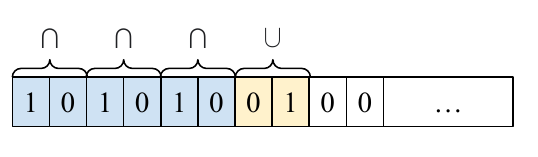}
  \caption{Example of bit-packing of 2-bit operators for a stack containing three set intersection operators followed by a single set union operator.}
  \label{fig:bitpack-op}
\end{figure}

For prefix evaluation, rather than a stack of Boolean values, one needs to
maintain a stack of operators. In this case, there are three possible values
that need to be encoded: union, intersection, and a null value. Thus, each
operator can be stored in 2 bits. By a similar argument, with a 32-bit unsigned
integer one can carry out \cref{alg:prefix} with as many as 16 operators on the
stack at a given time. In our implementation in OpenMC, the stack stores the
union operator as 01, the intersection operator as 10, and the null value as 00.
An example of the memory layout of a stack containing four operators is shown in
\cref{fig:bitpack-op}. In our tests using the ITER E-lite model, the operator
stack never reached a size greater than 8 during point containment queries,
confirming that a maximum stack size of 16 should be sufficient for nearly all
use cases.

\subsection{Infix Evaluation}

\noindent Postfix and prefix notation have several advantages over infix
notation, including the ability to enforce operator precedence without
parentheses and straightforward evaluation algorithms, as demonstrated by
\cref{alg:postfix} and \cref{alg:prefix}. This is why several popular MC
particle transport codes rely on postfix notation rather than infix. While we
are aware of no codes that use prefix notation, the previous section showed that
it can enable short-circuiting logic, which is not possible in postfix notation.
However, we saw that with prefix notation it is necessary to manage a stack of
2-bit operators. While infix notation does not lend itself to as simple an
evaluation algorithm as for postfix and prefix notations, we show here that it
enables short-circuiting logic while requiring no stack data structure to manage
during the evaluation of an expression.

\begin{figure}[ht]
  \centering
  \includegraphics[width=3in]{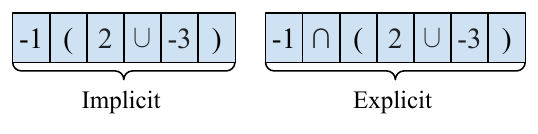}
  \caption{Memory representation (left-to-right) of the infix notation for the
  expression ``${-1} \cap ({+2} \cup {-3})$''.}
  \label{fig:infix-memory}
\end{figure}

Before presenting the algorithms for evaluating point containment using an infix
notation, there is a peculiarity of the CSG representation in MC particle
transport codes that must be considered. Namely, some codes (OpenMC, MCNP, and
Serpent, for example) have an input representation wherein intersection
operators are implicit; that is, the juxtaposition of two operands is
interpreted as the intersection between the operands. With that representation,
``${-1} \: {-2}$'' is the intersection of the negative half-spaces of two
surfaces. This gives us two choices as to how to store the notation in memory
and, correspondingly, how to evaluate point containment given the notation. If
intersection operators are not stored in memory, we refer to the representation
as being implicit; an example of this is shown in the left side of
\cref{fig:infix-memory}. If intersection operators \emph{are} stored in memory,
the representation is explicit (right side of \cref{fig:infix-memory}).

Note that one important difference between the infix algorithms and the
postfix/prefix algorithms is that the infix algorithm must keep track of
parentheses, which now appear as tokens. For both infix evaluation algorithms,
we assume that parentheses are always present to enforce precedence between
operators, which may require additional preprocessing at initialization. That
is, if the user has provided an expression ``${-1} \cap{2} \cup {-3}$'', rather
than internally relying on the assumed precedence of the $\cup$ and~$\cap$
operators, a pair of parentheses is always inserted (e.g., ``${-1} \cap ({2}
\cup {-3})$'').

The choice of an implicit or explicit memory representation has a direct impact
on the corresponding evaluation algorithm. In the evaluation algorithm for an
implicit representation, shown in \cref{alg:infix-implicit}, the algorithm
itself must deduce when an intersection operator is implied by the presence of
two successive operands. Explicitly storing intersection operators in the memory
representation simplifies the resulting algorithm. \Cref{alg:infix-explicit}
outlines the infix evaluation algorithm for an expression with intersection
operators explicitly represented. We see that while an implicit representation
minimizes memory use for the expressions, it does so at the expense of a more
complex evaluation algorithm.

Both infix evaluation algorithms utilize short-circuiting that depends on the
relative depth of parentheses as well as lazy evaluation of primitives. The
short-circuiting algorithm, which simply skips ahead to the appropriate closing
parenthesis in the notation, is given in \cref{alg:infix-short-circuit}.
Importantly, the use of short-circuiting logic is what allows the algorithm to
keep track of only a single Boolean value rather than needing to store a stack
of Boolean values. To understand this, note that the $\lor$ operator is defined
such that $x \lor y$ returns $x$ if $x$ is true and $y$ otherwise. So, when $x$
is true, the right operand does not need to be evaluated. When $x$ is false, the
result of the expression $x \lor y$ is $y$, meaning there is no need to store
the value of $x$. A similar argument applies to $x \land y$; the expression
short circuits when $x$ is false and otherwise returns $y$, meaning the value of
$x$ needn't be stored.

\begin{algorithm}
  \caption{Infix surface half-space evaluation with implicit intersections. Note
  that it is assumed that parentheses have been inserted to enforce precedence.}
  \label{alg:infix-implicit}
  \begin{algorithmic}
    \Require $T \equiv$ CSG expression in prefix notation
    \Require $\mathbf{r} \equiv$ point in Euclidean space \\
    \Function{ContainsInfixImplicit}{$T,\mathbf{r}$}
      \State $b \gets \text{TRUE}$ \Comment{Status variable}
      \State $d \gets 0$ \Comment{Parentheses depth}
      \State $i \gets 1$ \Comment{Index in $T$}
      \While{$i <$ \textsc{Size($T$)}}
        \State{$x \gets T[i]$} \Comment{Get next node}
        \If{$x$ is a primitive}
          \If{$b = $ TRUE}
            \State $b \gets$ \textsc{Evaluate}($x,\mathbf{r}$)
          \ElsIf{$d = 0$}
            \State \textbf{break}
          \EndIf
        \ElsIf{$x$ is $\cup$}
          \If{$b =$ FALSE}
            \State $b \gets$ TRUE \Comment{Reset return status}
          \ElsIf{$d$ is $0$}
            \State \textbf{break}
          \Else
            \State $d \gets d - 1$
            \State \textsc{ShortCircuitInfix}$(T, i)$
          \EndIf
        \ElsIf{$x$ is (}
          \If{$b =$ FALSE}
            \State \textsc{ShortCircuitInfix}$(T, i)$
          \Else
            \State $d \gets d + 1$
          \EndIf
        \ElsIf{$x$ is )}
          \State $d \gets d - 1$
        \EndIf
      \EndWhile
      \State \textbf{return} $b$
    \EndFunction
  \end{algorithmic}
\end{algorithm}

\begin{algorithm}
  \caption{Infix surface half-space evaluation with explicit intersections}
  \label{alg:infix-explicit}
  \begin{algorithmic}
    \Require $T \equiv$ CSG expression in prefix notation
    \Require $\mathbf{r} \equiv$ point in Euclidean space \\
    \Function{ContainsInfix}{$T,\mathbf{r}$}
      \State $b \gets$ TRUE \Comment{Boolean value to be returned}
      \State $d \gets 0$ \Comment{Parentheses depth}
      \State $i \gets 1$ \Comment{Index in $T$}
      \While{$i <$ \textsc{Size($T$)}}
        \State{$x \gets T[i]$} \Comment{Get next node}
        \If{$x$ is a primitive}
            \State $b \gets$ \textsc{Evaluate}($x,\mathbf{r}$)
        \ElsIf{($x$ is $\cup$ \textbf{and} $b=$ TRUE) \textbf{or}
                 ($x$ is $\cap$ \textbf{and} $b=$ FALSE)}
            \If {$d=0$}
                \State \textbf{break}
            \Else
                \State $d \gets d - 1$
                \State \textsc{ShortCircuitInfix}($T, i$)
            \EndIf
        \ElsIf{$x$ is (}
            \State $d \gets d + 1$
        \ElsIf{$x$ is )}
            \State $d \gets d - 1$
        \EndIf
      \EndWhile
      \State \textbf{return} $b$
    \EndFunction
  \end{algorithmic}
\end{algorithm}

\begin{algorithm}
  \caption{Short-circuiting algorithm used by both infix algorithms}
  \label{alg:infix-short-circuit}
  \begin{algorithmic}
    \Require $T \equiv$ CSG expression in prefix notation
    \Require $i \equiv$ Index in $T$ \\
    \Function{ShortCircuitInfix}{$T,i$}
        \State $d' \gets 1$
        \While{$d' > 0$} \Comment{Loop until end of parentheses}
          \State $i \gets i + 1$
          \State $x \gets T[i]$ \Comment{Get next node}
          \If{$x$ is (}
            \State $d' \gets d' + 1$
          \ElsIf{$x$ is )}
            \State $d' \gets d' - 1$
          \EndIf
        \EndWhile
    \EndFunction
  \end{algorithmic}
\end{algorithm}

%-------------------------------------------------------------------------------
\section{Results}
\label{sec:results}

\noindent To measure the performance of the point containment evaluation
algorithms based on postfix, prefix, and infix notations, all of them have been
implemented on separate branches of the OpenMC particle transport code.
Additionally, several variations on each algorithm were studied:
\begin{itemize}
  \item \textbf{Postfix}---OpenMC version 0.13.1~\citep{openmc-0131} was used as
  the baseline for postfix notation. However, in this version, complement
  operators are present in the postfix notation, which therefore necessitates
  extra logic in the evaluation algorithm. A separate branch was created that
  uses De Morgan's laws to eliminate complements while otherwise preserving the
  postfix notation. Then, a third postfix branch that relies on a bit-packed
  stack allows us to study the impact of a more efficient stack data structure.
  \item \textbf{Prefix}---Two variations of the prefix evaluation algorithm were
  tested. First, a branch was created that implements the prefix evaluation
  algorithm and utilizes an operator stack based on the \texttt{std::vector}
  type. A second branch improves on this by utilizing a bit-packed operator
  stack.
  \item \textbf{Infix}---Two variations of the infix evaluation algorithm were
  tested: one that relies on an implicit memory representation and another that
  relies on an explicit memory representation.
\end{itemize}
Thus, there were seven different cases altogether that were tested, a summary of
which is shown in \cref{tab:cases}. A code diff containing the changes to OpenMC
for each of the cases is available in the data supplement to this article.
\begin{table}[htb]
  \centering
  \caption{Summary of OpenMC branches implementing different point containment algorithms.}
  \label{tab:cases}
  \begin{tabular}{cC{2cm}C{2cm}C{2cm}}
    \toprule
    \textbf{Algorithm} & \textbf{Complement expanded?} & \textbf{Bit-packed stack?} & \textbf{Implicit intersection?} \\
    \midrule
    Postfix-1 & & & --- \\
    Postfix-2 & \checkmark & & --- \\
    Postfix-3 & \checkmark & \checkmark & --- \\
    \midrule
    Prefix-1 & \checkmark & & --- \\
    Prefix-2 & \checkmark & \checkmark & --- \\
    \midrule
    Infix-1 & \checkmark & --- & \checkmark \\
    Infix-2 & \checkmark & --- & \\
    \bottomrule
  \end{tabular}
\end{table}

In OpenMC, CSG expressions are marked as either \emph{simple} or \emph{complex}
at initialization. Simple expressions are defined to be those that involve only
set intersection operators. For some CSG models, nearly all solids are
represented with simple CSG expressions. Thus, OpenMC---and other codes---have
separate evaluation logic for simple expressions as an optimization of the
common case where it is straightforward to implement short-circuiting and lazy
evaluation of primitives regardless of the notation being used. In order to test
the efficacy of the algorithms presented here, then, we need a model that
contains many solids with complex CSG expressions. For that, we rely on a
detailed, realistic MCNP model of the ITER tokamak~\citep{holtkamp2007fed} called
E-lite~\citep{juarez2021nature}. Whereas previous analysis models of the ITER
tokamak only represented limited segments of the full
machine~\citep{leichtle2018fed}, the E-lite model covers the full machine,
thereby eliminating many previous modeling approximations. The E-lite model
contains over 300,000 solids that are defined using over half a million implicit
surfaces, making it---to our knowledge---one of the most complex radiation
transport models in existence at the time of writing.

\begin{figure}[ht]
  \centering
  \includegraphics[width=\columnwidth]{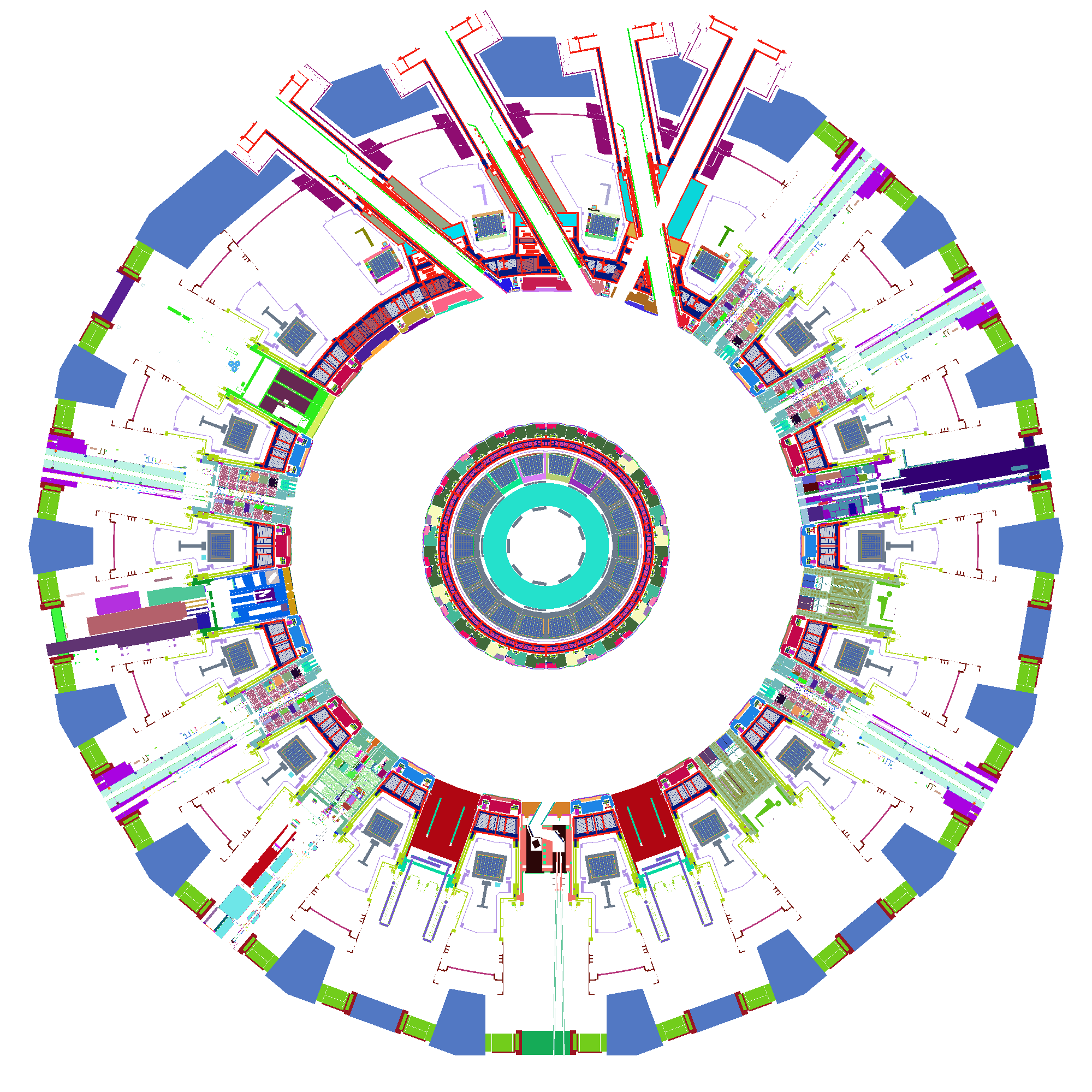}
  \caption{Horizontal ($x$-$y$) cross-sectional view of the ITER E-lite OpenMC model at an elevation of $z$=60 cm.}
  \label{fig:iter-xy}
\end{figure}

\begin{figure}[ht]
  \centering
  \includegraphics[width=3in]{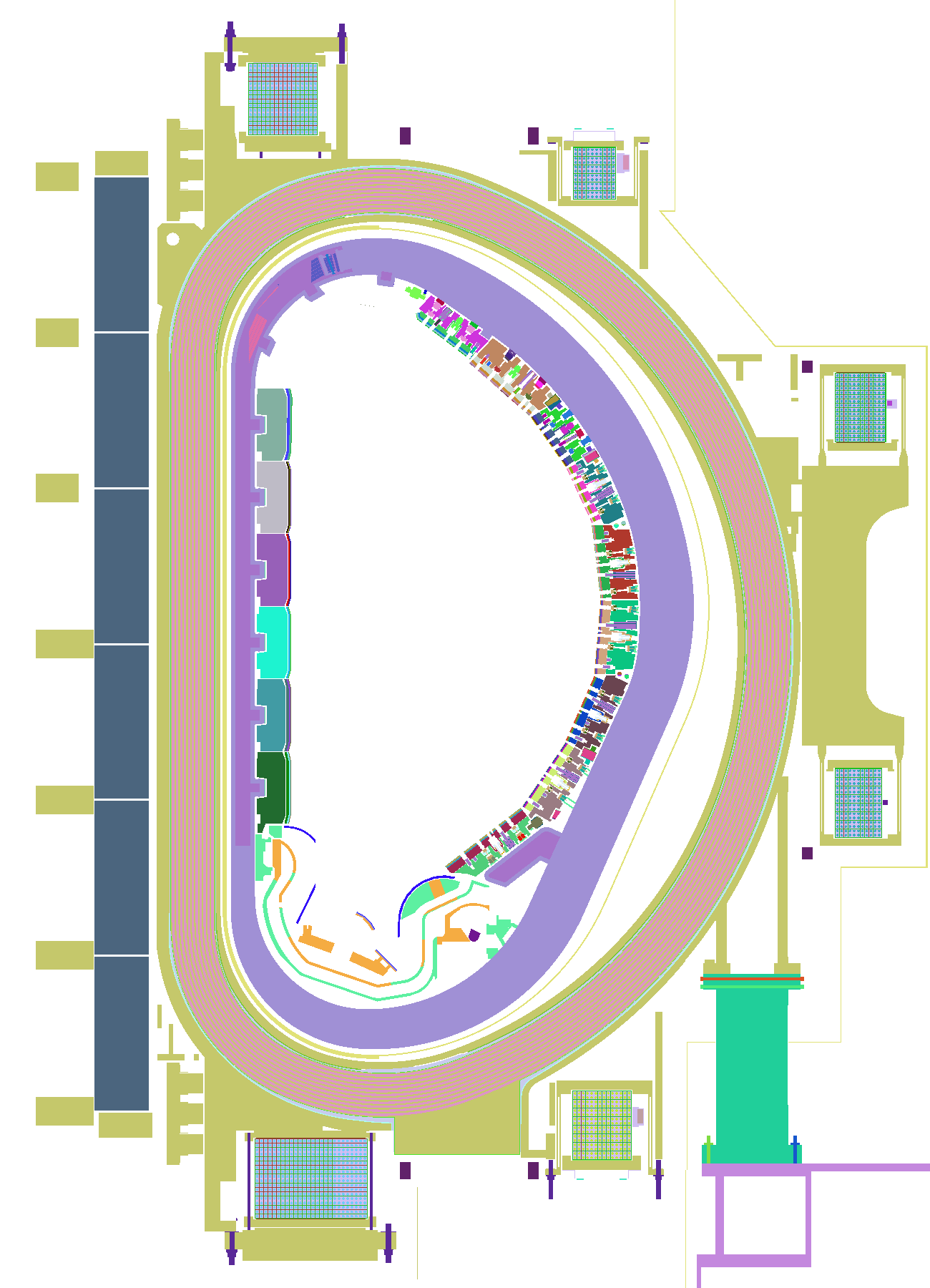}
  \caption{Vertical ($x$-$z$) cross-sectional view of the ITER E-lite OpenMC model.}
  \label{fig:iter-xz}
\end{figure}

In order to use the E-lite MCNP model with OpenMC, it was converted using an
automated MCNP model conversion utility~\citep{mcnpconverter}.
\Cref{fig:iter-xy,fig:iter-xz} show horizontal and vertical cross-sectional
views of the resulting OpenMC model that were generated using OpenMC's built-in
visualization capabilities. The model conversion utility seamlessly handles
conversion of implicit surfaces, solid bodies, and the material composition
definitions that are assigned to solid bodies. Some aspects of the model are not
handled, however, including the definition of the neutron source, which is
needed in order to carry out a neutron transport simulation. For the OpenMC
model, the neutron source definition was manually specified as the superposition
of 40 individual sources in $(r, \phi, z)$ to approximate the original source
definition in the MCNP model.

From here, we present our results in three parts. First, in \cref{sec:viz}, we
present performance results for the generation of a 2D slice visualization of
the E-lite model based on rasterization using each of the algorithms in
\cref{tab:cases}. Then, in \cref{sec:transport}, performance results are given
for neutron transport simulations of E-lite using each algorithm in
\cref{tab:cases}. Finally in \cref{sec:profile}, we show performance profiles of
OpenMC simulations of the E-lite model utilizing each algorithm to better
understand where performance improvements arise for the prefix and infix
notations.

\subsection{Rasterization Performance}
\label{sec:viz}

To isolate the performance impact of the point containment algorithms as much as
possible, our first integrated test is based on a function in OpenMC that
performs rasterization to produce a 2D slice visualization, such as those shown
in \cref{fig:iter-xy,fig:iter-xz}. The function iterates over a set of Cartesian
points in a given plane and, for each one, performs a search to determine what
solid body the point is contained in. For each point, the search proceeds by
iterating over all top-level solid bodies and performing a point containment
search on its associated CSG expression until one is found that contains the
point. Because nearly all the execution time is spent in point containment
searches, this function allows us to measure the performance impact of the
different algorithms without being diluted by other transport operations.

Each of the branches listed in \cref{tab:cases} was used to produce a single
high-resolution 2D slice visualization with 3600$\times$3600 pixels with a call
to the \texttt{openmc\_id\_map} C API function. Three iterations were performed
for each branch and we took the lowest time for each branch. The runs were
carried out on the Bebop cluster maintained by the Laboratory Computing Resource
Center at Argonne National Laboratory. Each test was run on one Intel Xeon
E5-2695v4 processor with 18 cores. The timing results are presented in
\cref{tab:raster}. The baseline version of OpenMC (Postfix-1) took
\SI{12.36}{\second} to generate the rasterized image. Interestingly, expanding
complement operators (Postfix-2) appears to slightly worsen the performance.
However, the use of a bit-packed stack for Boolean values results in a 50\%
improvement over the baseline. Adopting a prefix notation (Prefix-1), which
allows the evaluation algorithm to benefit from short-circuiting, performs even
better with an execution time 2.3 times faster than the baseline. Improving the
data structure for the operator stack (Prefix-2) improves the performance even
more, resulting in a 3$\times$ speedup over the baseline. Finally, the infix
evaluation algorithms perform the best of all, with improvements of 5.29$\times$
(implicit) and 6.11$\times$ (explicit) relative to the baseline.
\begin{table}[htb]
  \centering
  \caption{Timing results for producing a high-resolution raster plot using
  OpenMC with each point containment algorithm. Performance improvement is
  measured relative to Postfix-1.}
  \label{tab:raster}
  \begin{tabular}{ccC{2cm}}
    \toprule
    \textbf{Algorithm} & \textbf{Time [s]} & \textbf{Performance improvement} \\
    \midrule
    Postfix-1 & 12.36 & 1.00 \\
    Postfix-2 & 12.61 & 0.98 \\
    Postfix-3 & 8.32 & 1.49 \\
    \midrule
    Prefix-1 & 5.38 & 2.30 \\
    Prefix-2 & 4.17 & 2.97 \\
    \midrule
    Infix-1 & 2.34 & 5.29 \\
    Infix-2 & 2.02 & 6.11 \\
    \bottomrule
  \end{tabular}
\end{table}

\subsection{Transport Performance}
\label{sec:transport}

\noindent The results in \cref{sec:viz} show that using prefix or infix notation
results in significant performance improvements for point containment operations
relative to postfix notation. To demonstrate the impact in an actual neutron
transport simulation, we have carried out full simulations of the E-lite OpenMC
model using each algorithm from \cref{tab:cases}. Once again, all simulations
were run on one node of the Bebop cluster containing two Intel Xeon E5-2695v4
processors. Each OpenMC simulation used two MPI processes, 18 OpenMP threads per
process, 10,000 source particles per batch, and 10 total batches. OpenMC reports
the calculation throughput in terms of source particles simulated per second.
For each algorithm, we ran three simulations and took the best
result\footnote{We note that the results across the three independent runs were
very consistent in all cases, with less than 1\% difference in the measured
throughput.}. The performance results for the transport simulations are shown in
\cref{tab:transport}. The results follow the same trends as in \cref{tab:raster}
with a few exceptions. In this case, Postfix-2 performs slightly better than
Postfix-1, where for rasterization it performed slightly worse. For the
transport simulation results, it is expected that the overall performance
improvement would be slightly less than for rasterization since the simulation
involves many other operations. Nevertheless, utilization of point containment
algorithms based on prefix and infix notation offer up to 2.50$\times$ and
4.59$\times$ performance improvement, respectively, for the \emph{entire}
transport simulation of the E-lite model. The impressive performance results
here led to the inclusion of the algorithm based on infix notation with an
explicit memory representation in version 0.13.2 of OpenMC~\citep{openmc-0132}.
\begin{table}[htb]
  \centering
  \caption{Measured simulation throughput in source particles per second for
  neutron transport simulations of the E-lite model using OpenMC with each point
  containment algorithm. Performance improvement is measured relative to
  Postfix-1.}
  \label{tab:transport}
  \begin{tabular}{ccC{2cm}}
    \toprule
    \textbf{Algorithm} & \textbf{Rate [particles/s]} & \textbf{Performance improvement} \\
    \midrule
    Postfix-1 & 180.9 & 1.00 \\
    Postfix-2 & 187.3 & 1.04 \\
    Postfix-3 & 251.1 & 1.39 \\
    \midrule
    Prefix-1 & 328.0 & 1.81 \\
    Prefix-2 & 452.8 & 2.50 \\
    \midrule
    Infix-1 & 740.6 & 4.09 \\
    Infix-2 & 830.4 & 4.59 \\
    \bottomrule
  \end{tabular}
\end{table}

\subsection{Performance Profiles}
\label{sec:profile}

\begin{figure}[ht]
  \centering
  \includegraphics[width=3.5in]{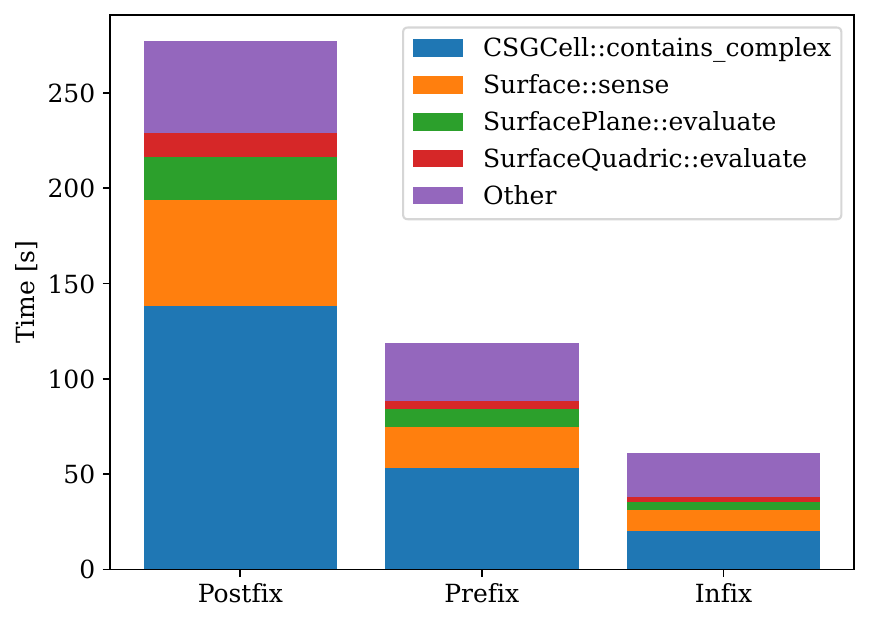}
  \caption{Time spent in selected functions during OpenMC execution of a neutron
    transport simulation of the ITER E-lite model based on each evaluation
    algorithm.}
  \label{fig:profile}
\end{figure}

\noindent During a particle transport simulation, many different operations must
be carried out including collision physics, geometry lookups, cross-section
(interaction probability) lookups, and tallies. To provide further context to
the results in \cref{sec:transport}, we have utilized perf (also known as
perf\_events), a Linux performance analysis tool, to profile the execution of
several short simulation (10 batches, each with 1000 particles) of the E-lite
model using algorithms Postfix-1, Prefix-2, and Infix-2. Each branch of OpenMC
was compiled with the CMake option \texttt{OPENMC\_ENABLE\_PROFILE} turned
on\footnote{This option ensures that the gcc compiler uses the
\texttt{-fno-omit-frame-pointer} flag which is necessary for producing accurate
profiles.}. Neutron cross sections from ENDF/B-VIII.0~\citep{brown2018nds} were
used in the simulation. The simulations were run on a laptop with an Intel Core
i7-1260P processor and used only one OpenMP thread. \Cref{fig:profile} shows the
time spent in the most expensive functions for each algorithm. The four
functions that are specifically called out are as follows:
\begin{enumerate}
  \item \texttt{CSGCell::contains\_complex} --- This function handles the
  evaluation logic for point containment on a complex CSG expression
  (\cref{alg:postfix}, \cref{alg:prefix}, or \cref{alg:infix-explicit}).
  \item \texttt{Surface::sense} --- This function evaluates an implicit
  surface inequality as in \cref{alg:evaluate}.
  \item \texttt{SurfacePlane::evaluate} --- This function is called by
  \texttt{Surface::sense} when it needs to evaluate the implicit surface
  equation for a plane.
  \item \texttt{SurfaceQuadric::evaluate} --- This function is called by
  \texttt{Surface::sense} when it needs to evaluate the implicit surface
  equation for a generic quadric (e.g., an ellipsoid or elliptic paraboloid).
\end{enumerate}
Several observations can be made on the results in \cref{fig:profile}. First, we
see that the use of prefix and infix notation significantly reduces the time
spent evaluating the CSG binary tree expressions (blue). In addition, the time
spent evaluating point containment on primitives (orange, green, and red) are
also reduced significantly thanks to the use of short-circuiting logic. Lastly,
the time spent in all other functions (purple) is also slightly reduced as it
also includes a number of functions that are called fewer times due to
short-circuiting.

%-------------------------------------------------------------------------------
\section{Conclusions}
\label{sec:conclusions}

\noindent In this article, we have presented several algorithms for evaluating
point containment in CSG trees with unbounded primitives. While the algorithms
are not specific to the representative of the primitives, we have drawn on
examples from Monte Carlo particle transport simulations, which often use CSG
trees based on implicit surface inequalities. Three basic algorithms were
presented, each based on a different notation of the CSG expression: postfix,
prefix, and infix. It was shown that prefix and infix allow short-circuiting
logic to be used, which reduces the number of primitives that must be checked
during point containment. Furthermore, we discussed storage requirements for
intermediate results during the execution of point containment algorithms. Both
postfix and prefix notation necessitate the use of a stack during evaluation,
whereas infix notation eliminates the need for a stack altogether. Techniques
for minimizing memory use and, correspondingly, dynamic memory allocation for
the stack were discussed for evaluations using the postfix and prefix notations.

To evaluate the performance of the various algorithms, two sets of tests were
carried out using the OpenMC Monte Carlo code along with several modified
branches of it. First, the execution time to generate a high-resolution
rasterized image of a 2D slice of a detailed CSG model of the ITER tokamak was
measured with each of the various algorithms. Use of both prefix and infix
notations offered significant speedup over the postfix notation that has
traditionally been used in MC particle transport codes, with infix resulting in
a 6$\times$ reduction in execution time for image generation relative to
postfix. For postfix and prefix notations, which require a stack of intermediate
results, the use of a bit-packed stack led to roughly 40\% improvement in
performance over the use of a traditional C++ data structure like
\texttt{std::vector}. Finally, for infix notation, we looked at the difference
between explicitly representing set intersection operators in the notation
versus representing them implicitly; the result demonstrates that the explicit
representation outperforms the implicit representation due to simplifications in
the algorithm.

In addition to the rasterization tests, we also measured the execution time of
neutron transport simulations of the ITER E-lite model using each of the various
algorithms. The results and performance improvements reveal the same trends as
for rasterization, although the addition of other operations in the transport
loop means that the performance improvements are slightly less---in this case, a
4.59$\times$ overall speedup using the infix notation relative to the baseline
postfix notation.

While the results that we have reported here are specifically for CPU
architectures, we expect that the performance gains will translate to other
architectures as well. Anecdotally, we have observed similar performance gains
for fusion-relevant problems with the use of infix notation on GPUs within a GPU
port of OpenMC based on OpenMP device offloading~\citep{openmcgpu}.

It is important to recognize that the improvements demonstrated for the ITER
E-lite model are not representative of all MC transport simulations. As noted,
many MC codes are already optimized for CSG expressions involving only set
intersection operators. Many CSG models for particle transport simulations have
a high fraction of solids that can be represented with such expressions, and
thus we would not expect to realize any performance benefit for these models.
However, for geometrically complex models, which occur quite commonly for fusion
energy---and possibly other---applications, significant performance benefit can
be expected from the algorithmic advancements presented herein.

The scope of this paper has been limited solely to a single
operation---evaluating point containment on a CSG tree. Of course, there are
many other techniques one can use to accelerate spatial searches. In
\cref{sec:viz}, we described how the search for the solid body containing a
given point in OpenMC involves a linear search over all top-level solid bodies
in the model. Use of a bounding volume hierarchy or other spatial acceleration
data structure / algorithm can also greatly reduce the time needed to perform
these spatial searches. Other techniques to simplify and/or reorder the
operations in the CSG binary expression tree could similarly improve
performance.

\section*{Data Availability}

The data that support the findings of this study are openly available at the
following DOI:
\href{https://doi.org/10.5281/zenodo.10045317}{10.5281/zenodo.10045317}.

\section*{CRediT Author Statement}
\textbf{Paul K. Romano}: Conceptualization, Investigation, Visualization,
Writing --- Original Draft, Writing --- Review \& Editing, Supervision, Project
administration, Funding acquisition. \textbf{Patrick A. Myers}: Methodology,
Software, Investigation, Writing --- Original Draft. \textbf{Seth R. Johnson}:
Software, Writing --- Review \& Editing. \textbf{Alja\u{z} Kol\u{s}ek}:
Software, Writing --- Review \& Editing. \textbf{Patrick C. Shriwise}:
Conceptualization, Writing --- Review \& Editing.

\section*{Acknowledgments}
This work is supported by the U.S. Department of Energy Office of Fusion Energy
Sciences under award number DE-SC0022033. We gratefully acknowledge the
computing resources provided on Bebop, a high-performance computing cluster
operated by the Laboratory Computing Resource Center at Argonne National
Laboratory. This work was carried out using an adaption of the E-lite model
which was developed as a collaborative effort between: AMEC Co (International),
CCFE (UK), ENEA Frascati (Italy), FDS Team of INEST (PRC), ITER Organization
(France), QST (Japan), KIT (Germany), UNED (Spain), University of
Wisconsin-Madison (USA), F4E (Europe). The views and opinions expressed herein
do not necessarily relect those of the ITER Organization.

\bibliographystyle{elsarticle-num}
\bibliography{references}

\clearpage
\vspace*{\fill}
\noindent\fbox{%
  \parbox{\textwidth}{%
    The submitted manuscript has been created by UChicago Argonne, LLC, Operator
    of Argonne \mbox{National} Laboratory (``Argonne'').  Argonne, a
    U.S. Department of Energy Office of Science laboratory, is operated
    under Contract No. \mbox{DE-AC02-06CH11357}.  The U.S. Government retains for
    itself, and others acting on its behalf, a paid-up nonexclusive, irrevocable
    worldwide license in said article to reproduce, prepare derivative works,
    distribute copies to the public, and perform publicly and display publicly,
    by or on behalf of the Government. The Department of Energy will provide
    public access to these results of federally sponsored research in accordance
    with the DOE Public Access
    Plan. \url{https://energy.gov/downloads/doe-public-access-plan}
  }%
}
\vspace*{\fill}

\end{document}